\documentclass[apjl]{emulateapj}
\slugcomment{{\sc Accepted to ApJL:} September 21, 2012} 
\usepackage{graphicx,amsmath,natbib}

\begin{document}
\shorttitle{Electron Acceleration in Solar Coronal Jets}
\title{\sc Particle-In-Cell Simulation of Electron Acceleration in Solar Coronal Jets}

\shortauthors{Baumann \& Nordlund}
\author{G. Baumann and \AA. Nordlund}
\affil{Niels Bohr Institute, Juliane Maries Vej 30, DK-2100 K\o benhavn \O , Denmark}
\email{gbaumann@nbi.ku.dk}
% $Id: ms_full.tex,v 1.12 2012/09/21 09:31:34 gbaumann Exp $
%%%%%%%%%%%%%%%%%%%%%%%%%%%%%%%%%%%%%%%%%%%%%%%%%%%%
\newcommand{\fig}[1]{Fig.\ \ref{fig:#1}} % Fig. reference
\newcommand{\Fig}[1]{Figure \ref{fig:#1}} % Figure reference
\newcommand{\bea}[1]{\begin{eqnarray}\label{eq:#1}}
\newcommand{\eea}{\end{eqnarray}}
\newcommand{\beq}[1]{\begin{equation}\label{eq:#1}}
\newcommand{\eeq}{\end{equation}}
\newcommand{\Eq}[1]{Equation~\ref{eq:#1}}
\newcommand{\Eqs}[2]{Equations~\ref{eq:#1}--\ref{eq:#2}}
\newcommand{\Section}[1]{Section~\ref{sec:#1}}
\newcommand{\Table}[1]{Table~\ref{table:#1}}
\newcommand{\ddt}[1]{\frac{\partial #1}{\partial t}}
\newcommand{\ddx}[1]{\frac{\partial #1}{\partial x}}
\newcommand{\ddy}[1]{\frac{\partial #1}{\partial y}}
\newcommand{\ddz}[1]{\frac{\partial #1}{\partial z}}
\newcommand{\ddi}[1]{\frac{\partial #1}{\partial r_i}}
\newcommand{\ddj}[1]{\frac{\partial #1}{\partial r_j}}
\newcommand{\DDt}[1]{\frac{D #1}{ D t}}

\providecommand{\e}[1]{\ensuremath{\times 10^{#1}}}
\renewcommand{\div}{\nabla\cdot}
\newcommand{\curl}{\nabla\times}
\newcommand{\grad}[1]{{\nabla #1}}
\newcommand{\uu}{\mathbf{u}}
\renewcommand{\gg}{\mathbf{g}}
\newcommand{\pp}{\mathbf{p}}
\newcommand{\JJ}{\mathbf{J}}
\newcommand{\BB}{\mathbf{B}}
\newcommand{\EE}{\mathbf{E}}
\newcommand{\ux}{u_{\rm x}}
\newcommand{\uy}{u_{\rm y}}
\newcommand{\uz}{u_{\rm z}}
\newcommand{\gz}{g_{\rm z}}
\newcommand{\Hp}{H_{\rm P}}
\newcommand{\rr}{\mathbf{r}}
\newcommand{\Om}{\mathbf{\Omega}}
\newcommand{\Frad}{\mathbf{F}_{\rm rad}}
\newcommand{\Qrad}{Q_{\rm rad}}
\newcommand{\Qvisc}{Q_{\rm visc}}
\newcommand{\Qjoule}{Q_{\rm Joule}}
\newcommand{\Tvisc}{\mathbf{\tau}_{\rm visc}}
\newcommand{\Tij}{\tau_{ij}}
\newcommand{\sij}{s_{ij}}
\newcommand{\Bv}{B_{\nu}}
\newcommand{\Sv}{S_{\nu}}
\newcommand{\Iv}{I_{\nu}}
\newcommand{\tauv}{\tau_{\nu}}
\newcommand{\kv}{\kappa_{\nu}}
\newcommand{\Ekin}{E_{\rm kin}}
\newcommand{\Etherm}{E}
\newcommand{\Fconv}{\mathbf{F}_{\rm conv}}
\newcommand{\Fvisc}{\mathbf{F}_{\rm visc}}
\newcommand{\Fkin}{\mathbf{F}_{\rm kin}}
\newcommand{\half}{\frac{1}{2}}
\newcommand{\lbc}{{\rm bot}}

%%%%%%%%%%%%%%%%%%%%%%%%%%%%%%%%%%%%%%%%%%%%%%%%%%%%%%%%%%%%%%%%%%%%%%%%%%%%%%%%%%%%%%
\begin{abstract}
We investigate electron acceleration resulting from 3D magnetic reconnection between an
emerging, twisted magnetic flux rope and a pre-existing weak, open magnetic
field.  We first follow the rise of an unstable, twisted flux tube with a
resistive MHD simulation where the numerical resolution is enhanced by using
fixed mesh refinement. As in previous MHD investigations of similar situations,
the rise of the flux tube into the pre-existing inclined coronal magnetic field
results in the formation of a solar coronal jet.  A snapshot of the MHD model
is then used as an initial and boundary condition for a particle-in-cell
simulation, using up to half a billion cells and over 20 billion charged particles.
Particle acceleration occurs mainly in the reconnection current sheet, with
accelerated electrons displaying a power law in the energy probability distribution with an index
of around --1.5. The main acceleration mechanism is a systematic electric field,
striving to maintaining the electric current in the current sheet against
losses caused by electrons not being able to stay in the current sheet for more
than a few seconds at a time.
\end{abstract}

\subjectheadings{acceleration of particles --- Sun: corona --- Sun: magnetic topology}

%%%%%%%%%%%%%%%%%%%%%%%%%%%%%%%%%%%%%%%%%%%%%%%%%%%%%%%%%%%%%%%%%%%%%%%%%%%%%%%%%%%%%%
\section{\uppercase{Introduction}}
\label{sec:introduction}

Solar jets have been shown to be triggered by magnetic reconnection, similarly
to solar flares, while their released energy is much below what is set free in
a medium-sized flare event, and the timescales are usually shorter. Nevertheless
their high frequency of occurrence makes them a significant contributor to the
solar ejecta, particularly the solar wind originating from coronal holes.
Solar jets feature upflow velocities of more than 150\,km\,s$^{-1}$
\citep{2007PASJ...59S.771S,2008A&A...481L..57C}
and are observed at EUV down to X-ray wavelengths primarily in coronal holes
\citep{2007PASJ...59S.757K},
but also in active regions
\citep{2008A&A...481L..57C}.
%Observational data of solar jets has mainly been provided by YOHKOH
%\citep{1992PASJ...44L.173S,1996PASJ...48..123S}, SOHO
%\citep{1997Natur.386..811I}, HINODE \citep{2008A&A...481L..57C}, TRACE
%\citep{1999SoPh..190..167A} and SDO \citep{2011A&A...534A..62S}. 
Using RHESSI data, \citet{2007ApJ...663L.109K,2011ApJ...742...82K} have further investigated the electron impact regions of solar jets as a result of interchange reconnection. 

There have been several studies in the past employing fully 3D kinetic
models to study particle acceleration. Two approaches are most
popular: particle-in-cell (PIC) simulations and test particle simulations.
Their main difference is the back-reaction from the particles onto the fields,
which is only taken into account in the former method, while it is assumed to
be negligible in the latter one, justified by using a low number of test
particles for such simulations. These kind of simulations provide much information on
particle trajectories and favored locations of particle acceleration \citep{2005ApJ...620L..59T,2006A&A...449..749T,2005A&A...436.1103D,2006ApJ...640L..99D,2008A&A...491..289D}.
Nevertheless there are severe limitations to test particle simulations. These together with their consequences have in detail been discussed in
\citet{2010A&A...511A..73R}. On the other hand, the approach using
self-consistently evolving fields such as in PIC codes is subject to
several resolution constraints, reducing the possible physical
box size that can be simulated to far below length scales of solar jets. In
order to bypass these limitations, we made use of modifications
of the constants of nature. Thereby we are able to present results from fully
3D PIC simulations of a
solar jet, using essentially the same initial setup as in
\citet{2010A&A...511A..73R}, but with self-consistently evolving fields. Such modifications have previously to some extent been used by \citet{2006Natur.443..553D} and \citet{2009JPlPh..75..619S}. A comparison of simulations using different modifications exceeds the scope of this Letter. We therefore refer to \citet{2012arXiv1204.4947B} for a description and analysis of the physical consequences of modifications. For the present study we use a choice of modifications as close as we can get to reality with the currently available computational resources. 

\Section{simulations} provides an overview of the experiment and describes the MHD and PIC simulations as well as their intercoupling. In \Section{results} the results are presented
and discussed. Finally, in \Section{conclusions} conclusions are drawn.

%%%%%%%%%%%%%%%%%%%%%%%%%%%%%%%%%%%%%%%%%%%%%%%%%%%%%%%%%%%%%%%%%%%%%%%%%%%%%%%%%%%%%%
\section{\uppercase{Simulations}}
\label{sec:simulations}
\begin{figure}[ht]
    \centering
        \includegraphics[width=0.8\linewidth]{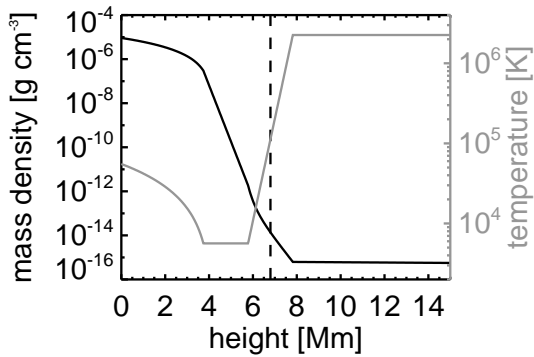}
    \caption{1D atmospheric profile for the MHD simulations. The dashed vertical line shows the lower cut in height for the sub-domain used for
    the PIC simulations.}
    \label{fig:density_temperature_profile}
\end{figure}

The solar jet experiment at hand starts out with a fully 3D resistive compressible MHD
simulation of a twisted emerging flux rope, initially positioned
1.7\,Mm below the photosphere, similar to the setup used by
\citet{2008ApJ...673L.211M}. A constant magnetic field of 3.3\,G
is imposed on the entire computational box, inclined 65$^{\circ}$ in the $yz$-plane. The maximum magnetic field strength
of the flux rope is slightly higher than 1000\,G and hence much larger than
the background magnetic field. The atmosphere is initially in hydrostatic
equilibrium with a 1D atmospheric profile similar to the one used in
\citet{2005ApJ...635.1299A}: the sub-photospheric temperature at the bottom is 5.5\e{4}\,K,
with a maximum mass
density $\rho$ of about 9\e{-6}\,g\,cm$^{-3}$ at a depth of 3.7\,Mm
below the surface.
The ``chromosphere'' has a constant temperature of around 5600\,K and the
corona starts out with $T =\,2.2\e{6}$\,K and $\rho = 6\e{-16}$\,g\,cm$^{-3}$, as
illustrated in Figure\,\ref{fig:density_temperature_profile}.

The simulations are performed using the \emph{Stagger} MHD code, as in
\citet{2008ApJ...673L.211M}, assuming an ideal gas law and neither taking heat
conduction nor radiative cooling into account. Viscosity and resistivity are
locally defined, depending mainly on the velocity divergence, in order to provide
a suitable (to maintain code stability), but minimal amount of dissipation. A mesh with 512$^{3}$ cells covers a box with physical extents of
$33.8\times38.1\times32.5$\,Mm.  The mesh is non-uniform in all directions,
with a minimal
mesh spacing of ($x_{min}$, $y_{min}$, $z_{min}$)=(0.034, 0.034, 0.030)\,Mm
around the reconnection region, and with a mesh spacing less than 10\% larger
than that in a region of size $12.5 \times 10.7 \times 5.1$\,Mm.
The $z$-coordinate is the direction normal to the solar surface.

Despite the slightly denser corona compared to the previously performed
simulations by  \citet{2008ApJ...673L.211M}, the overall evolution of the
experiment is the same: the twisted flux tube is made buoyantly unstable by
applying a density perturbation at its center, which causes it
to rise up into the much rarer corona, against the Lorentz force from the
bending tube, while expanding as described in \citet{2004A&A...426.1047A}.
Above the photosphere, the expansions of the magnetic flux rope, now due to the
high magnetic pressure, continue rapidly as more and more flux reaches
coronal heights. At the same time the corona is locally pushed upward where
plasma as well as magnetic flux emerges from below \citep[cf.][]{2005ApJ...635.1299A}. The
interaction between the two magnetic field domains defined by the initial coronal magnetic field and
the flux rope leads to a destabilization of the field configuration and causes
the formation of a thin dome-shaped current sheet where the magnetic field
lines are most inclined relative to each other. As in \citet{2008ApJ...673L.211M}, most
magnetic field lines of the two domains end up being almost anti-parallel at their
first encounter, which makes their interaction maximally powerful. The sheet is
subject to ohmic dissipation, causing it to reach temperatures as high as about
8\e{6}\,K. Reconnection gradually occurs between various field lines along the
twisted tube and the ambient coronal magnetic field. Due to this restructuring
of the magnetic field, several distinct flux domains form in the corona, as
shown in \citet[][see their Figure\,2]{2008ApJ...673L.211M}. A hot plasma jet pair
emerges from the high temperature and gas pressure region of the reconnection
region, propagating sideward together
with the expelled reconnected magnetic field lines. The plasma in the
jets is fed to the reconnection site from both sides of the current sheet; the
region below supplies dense and cold photospheric plasma, while the plasma
coming from above is much hotter and rarer coronal plasma.

After a large fraction of the dense plasma enclosed in the flux rope has
been reconnected and ejected in the form of plasma jets, as well as drained off along the magnetic field lines due
to gravity acting on the heavy sub-photospheric gas, we initialize
the PIC experiment with a cut-out of size 22 $\times$ 22
$\times$ 22\,Mm from the MHD data set. The reconnection process
continues in the PIC simulation expelling coronal low-density plasma and reconnecting field lines
from both connectivity domains.

The PIC simulations are performed using the \emph{Photon-Plasma} code
\citep{Haugboelle:2005,Hededal:2005b}, which solves the Maxwell equations
together with the relativistic equation of motion for charged particles. We fix
the magnetic fields at the boundaries to the values given by the MHD data set,
and leave the
boundaries open for particles to exit and enter \citep{Haugbolleetal2012}. To
initialize the electric field in the PIC simulation, only the advective
electric field ($-\textbf{u} \times \textbf{B}$, where $\textbf{u}$ is the bulk
speed) is passed on. The particles are initially given a random thermal
velocity drawn from a Maxwellian distribution, plus the bulk velocity from the
MHD simulation. Electron velocities consist of additionally the velocity due to the initial electric current:
\begin{equation}
v_{J} = \frac{1}{\mu_0 q n}  (\nabla \times \mathbf{B}),
\label{equ:driftvelo}
\end{equation}
where the magnetic field $\mathbf{B}$ and the density $n$ are provided by the
MHD snapshot data set. The jet velocity of this specific MHD data set is at that
point in time on the order of 400--800\,km\,s$^{-1}$. These jets are
dominated by the thermal motion in this high-temperature and low magnetic field
region.

We conducted several PIC runs with grid dimensions of 400$^3$ and 800$^3$ cells
on a uniform grid with cell sizes of 55\,km and 27.5\,km respectively, in
each case with 20 particles per species (protons and electrons) per cell, simulating up to 7.5 solar seconds.

\begin{figure*}[ht]
    \centering
\includegraphics[width=0.495\linewidth]{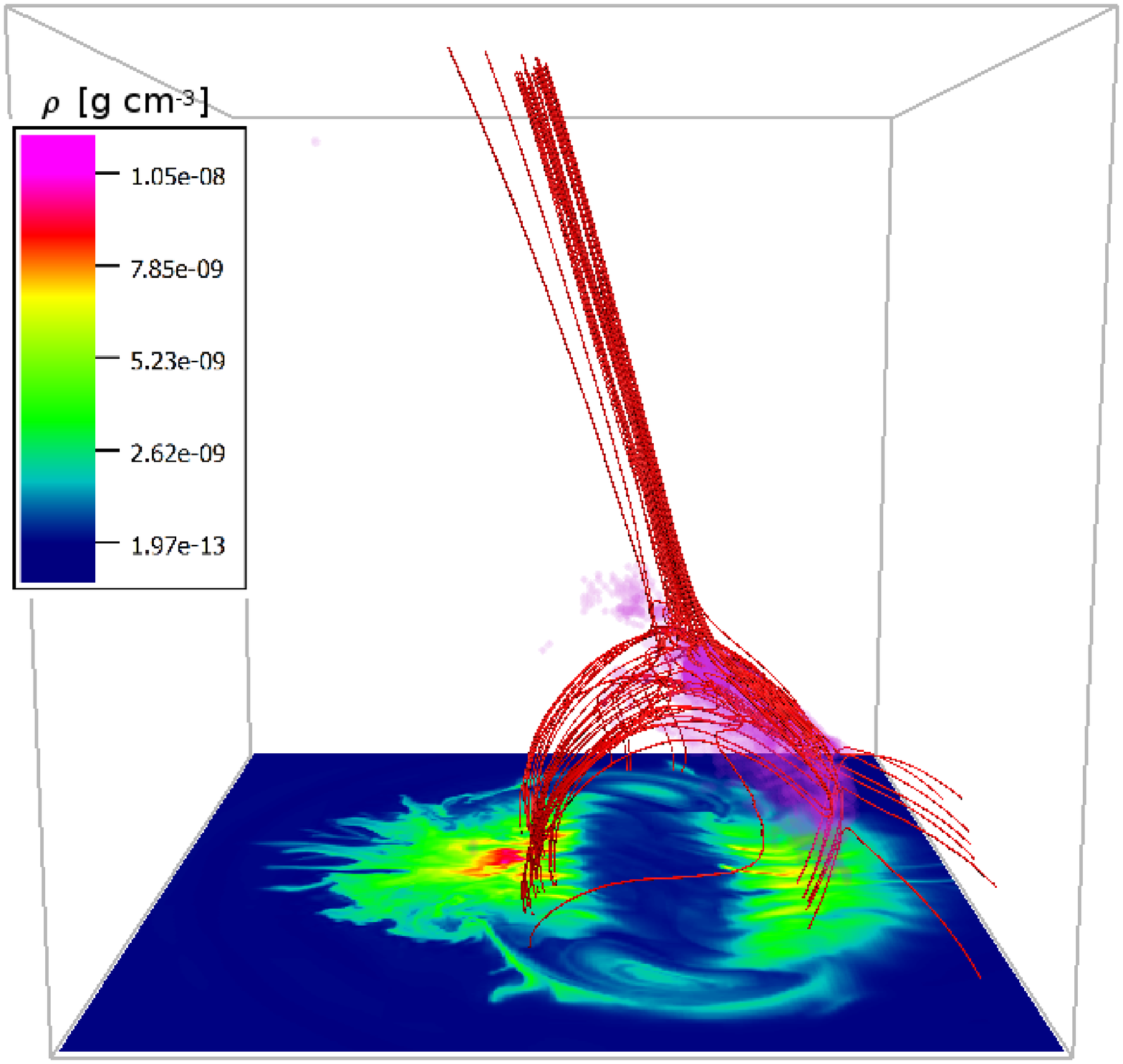}
\includegraphics[width=0.495\linewidth]{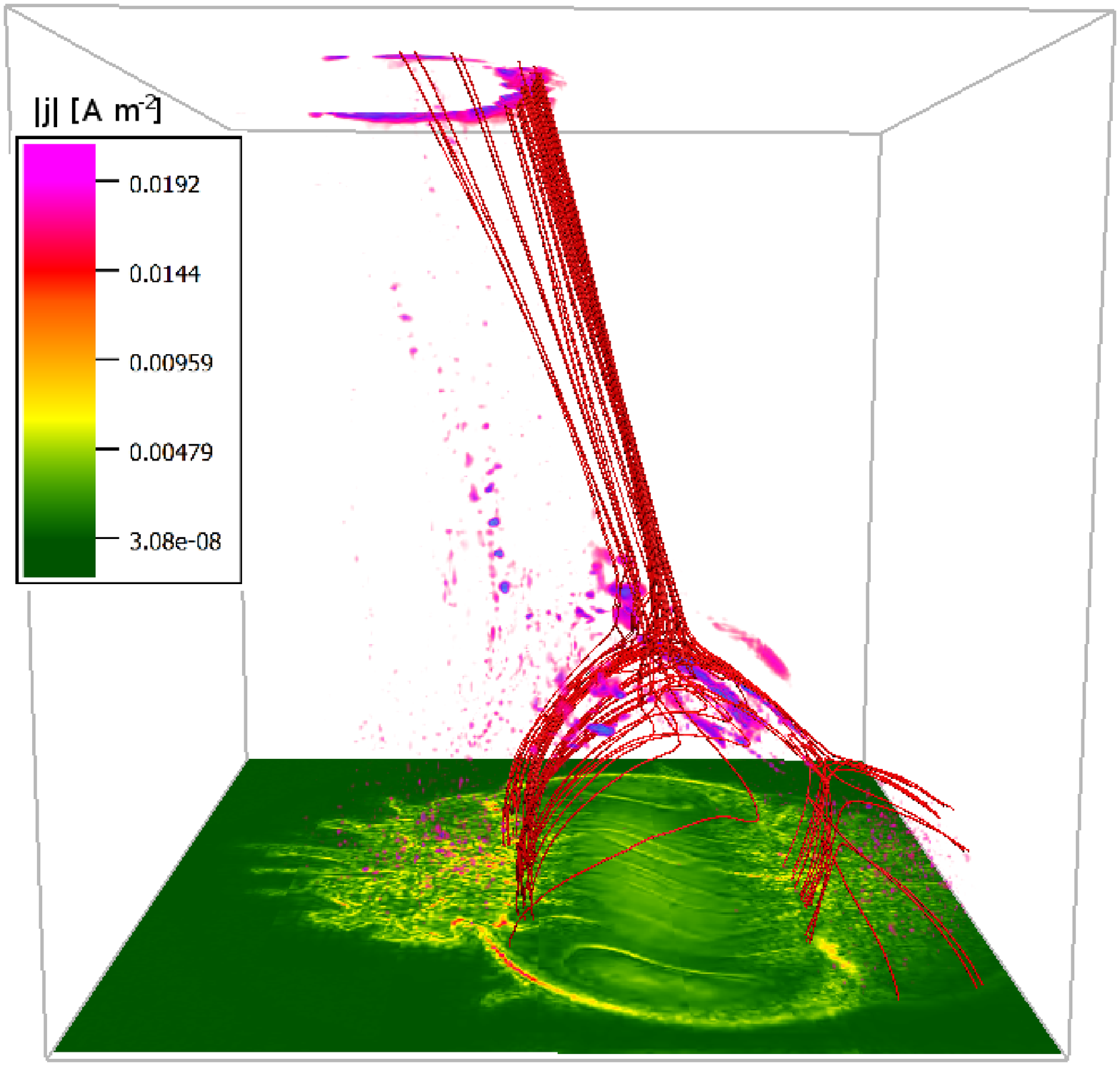}
    \caption{Magnetic field lines (red/black) with (a) the
    diffusive electric field $\mathbf{E_{\textrm{res}}} = \eta \mathbf{j}$ in the
    PIC cut-out of the MHD snapshot data set (purple/gray volume) and the charge
    density plane at the bottom of the box---note the low density inside the flux rope, as explained in the text---and (b) the $E_{\parallel}$ field
    0.5\,s after start of the PIC simulation in run 400$^{3}$ (purple/gray),
    together with the current density plane at the bottom of the box.}
    \label{fig:Eres_eb_3D}
\end{figure*}

To minimize computational constraints, the MHD snapshot is cut at 1.1\,Mm below
the bottom of the corona, hence in the transition region, shown as a vertical dashed line
in Figure\,\ref{fig:density_temperature_profile}. This limits the density span
to a factor of 4\e{4}, small compared to the span of about 2\e{10}
covered by the MHD simulation, but still large for a PIC simulation. Additionally, the temperature is reduced by a factor of four in order to avoid drowning the high-energy tail in the Maxwellian distribution.

To make the plasma micro-scales marginally resolvable, the charge per particle is reduced, and to ease the time step
constraint from the propagation of electro-magnetic waves the speed of light
is reduced, as explained in \citet{2012arXiv1204.4947B}. The electron skin depth in the current sheet is resolved
with about 5--10 grid cells. The electron gyroradius varies between a fraction
of a cell in the flux tube interior to many cells inside the current sheet---because
of the near cancellation of oppositely directed magnetic fields there are
several dynamically evolving null points inside the current sheet.

In addition to these stratified atmosphere simulation runs, control runs with a constant density of 3\e{-15}\,g\,cm$^{-3}$ were also performed. Both types
of runs show similar overall results.

%%%%%%%%%%%%%%%%%%%%%%%%%%%%%%%%%%%%%%%%%%%%%%%%%%%%%%%%%%%%%%%%%%%%%%%%%%%%%%%%
\section{\uppercase{Results and Discussion}}
\label{sec:results}

During the MHD flux emergence simulation a diffusive electric field
$\mathbf{E_{\textrm{res}}} = \eta \mathbf{j}$, where $\eta$ is the resistivity and $\mathbf{j}$
the electric current density, builds up in the reconnection region, approximately
cospatial with the current sheet. The parallel (in relation to the magnetic
field) electric field $E_{\parallel}$ is part of the diffusive (non-ideal) electric
field, and provides information on the rate of reconnection as well as on
favored regions for particle acceleration. The diffusive component of the
electric field is, unlike the advective electric field ($\mathbf{-u} \times
\mathbf{B}$), not inherited by the PIC code, but builds up self-consistently.
Figure\,\ref{fig:Eres_eb_3D} compares the location of the diffusive electric
field component for the chosen snapshot of the MHD simulations with
the $E_{\parallel}$ field of the PIC simulation 0.5\,s after start.
The parallel electric fields reach in general much higher magnitudes in the MHD simulations compared to the PIC simulations.
%The purple volumes cover values of about 0.16 -- 9142\,V\,m$^{-1}$ for the resistive MHD electric field and 143 -- 1071\,V\,m$^{-1}$, which is rescaling the modifications we applied around 0.6 -- 4.9\,mV m$^{-1}$ for the PIC parallel electric field, where the higher ends are the peak values of the entire computational box.

\begin{figure}[ht]
  \centering
  \includegraphics[width=\linewidth]{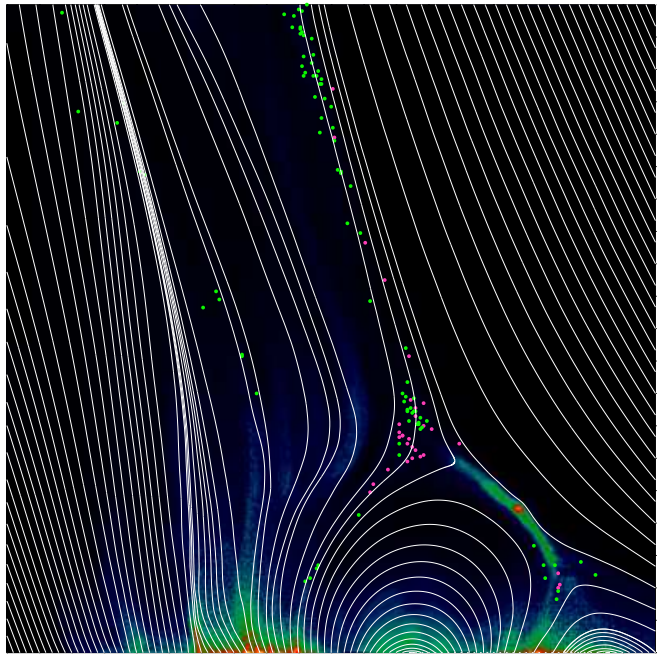}
  \caption{Electrons that win energy over a time
  interval of 2.5\,s, about 4.5\,s after the start of the 400$^3$   simulation run together with the magnetic field lines (white) and
  the electric current density as contour plot (blue--green--red/black--gray--white for increasing
  current density) in a $yz$-plane. Particles with velocities directed upward are colored
  purple/light gray, while downward moving electrons are colored green/gray.}
\label{fig:particle_pos_color.eps}
\end{figure}

$E_{\parallel}$ is the most efficient particle accelerator, since its
force acts on the particles without being affected by the perpendicular
particle gyromotion. Its maximum in the PIC simulations is located inside the current sheet, equivalent
to the diffusive electric field $E_{\textrm{res}}$ in the MHD simulation. Accelerated
electrons (see Figure\,\ref{fig:particle_pos_color.eps}) are located in the plasma outflow parts of
the reconnection region. The electron bulk velocity in the jet is on the order of 2000\,km\,s$^{-1}$. The proton bulk jet flow on the other hand is only about 270\,km\,s$^{-1}$, which difference to the electron bulk speed defines the electric current required by the magnetic field configuration.
%Sound and Alfv\'en speed in the current sheet area are only on the order of 20\,km\,s$^{-1}$.
The lower plane of the PIC simulation visualization in
Figure\,\ref{fig:Eres_eb_3D} (right) shows the electric current density. At the bottom
center of this figure resides the flux rope, whose twisted field lines are
indicated by the strongest electric current pattern. Additionally, to the left of the flux rope signature, the current sheet features a turbulent structure. This
is the result of the plasma transport caused by reconnection. In addition, we observe fast up and down flowing plasma, as can be seen in
Figure\,\ref{fig:particle_pos_color.eps}, in which upward moving electrons are
shown in purple/light gray and downward moving electrons are shown in green/gray.

\begin{figure*}[ht]
    \centering
    \includegraphics[width=0.83\linewidth]{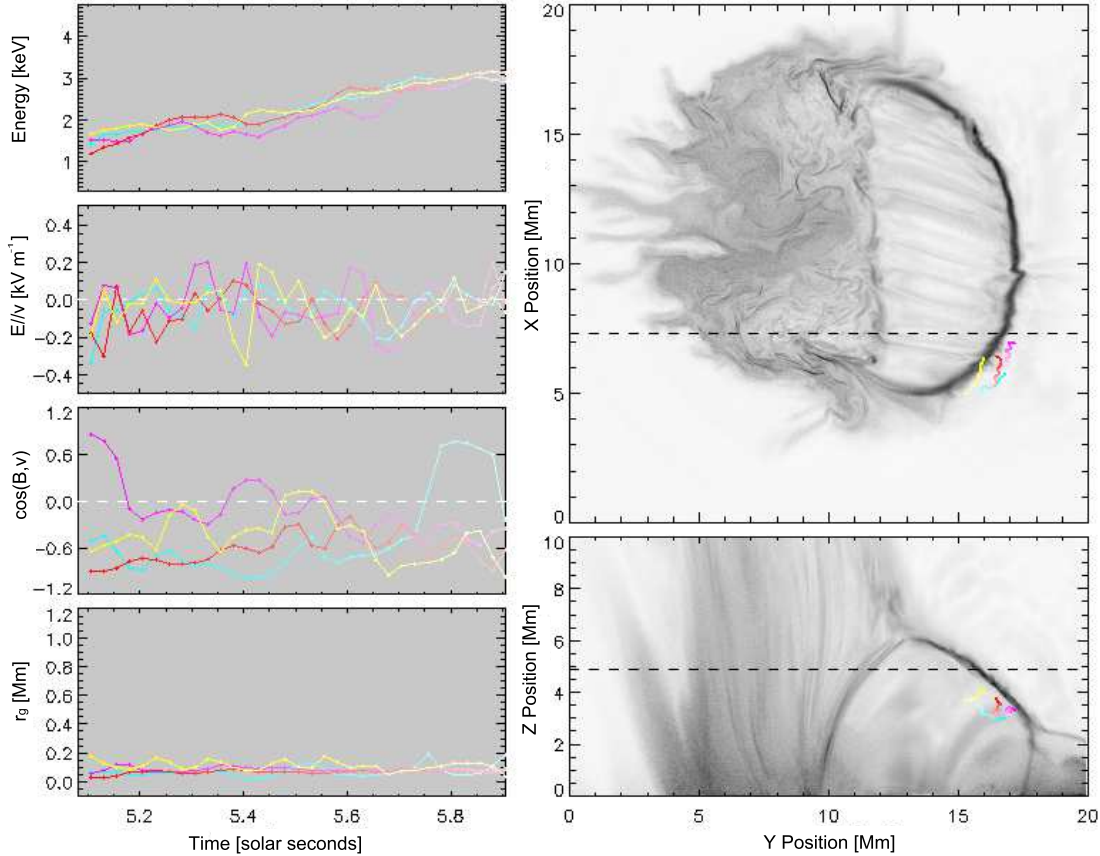}
    \caption{Four random electrons traced in run 800$^3$ during 1\,s. Their projected positions are plotted in the slices to the right,
    together with the electric current density (raised to the power 0.5 to
    enhance fine structures) in $yz$- and $yx$-planes. The black dashed
    lines in the right images show the cut in the respective direction for
    the other image. Additionally, the gyroradius $r_g$ of the
    particles, their cosine of the pitch angle $\cos(B,v)$, and the electric
    field in the direction of motion, $E$$\parallel$$v$, are plotted. }
    \label{fig:trace_plot_win.eps}
\end{figure*}

By tracing particles that win energy over a time period of a second, it
becomes clear, see Figure\,\ref{fig:trace_plot_win.eps}, that the acceleration
mechanism is a systematic DC electric field, particularly present in the
proximity of the current sheet, and mainly directed parallel to the magnetic
field. However, since the magnetic field is very weak in the current sheet area
particles are almost decoupled from the magnetic field lines, leading to pitch angle fluctuations seen in the third panel to the left in
Figure\,\ref{fig:trace_plot_win.eps}.

We only write out every 25th particle data, hence in Figures\,\ref{fig:particle_pos_color.eps} and \ref{fig:trace_plot_win.eps} one particle represents in reality 25 of its type.
Note that in Figure\,\ref{fig:trace_plot_win.eps} the electric field
fluctuations in the second panel on the left are to a large extent Monte Carlo
noise, due to low numbers of particles per cell. This has been verified with a
control experiment using
twice the number of particles per cell. Further, the background electric
current density image is not exactly at the particle positions, which results
in a slight projection offset.

Only a limited number of particles happen to be close enough to the current
sheet to be efficiently accelerated by $E_{\parallel}$, thus being able to
contribute to the electric current required by the magnetic field topology. After
experiencing a certain amount of acceleration, they get expelled from the
current sheet, as they follow magnetic field lines which leave the current sheet
region. The particles lost from the current sheet are replaced by new particles,
which again need to be accelerated. Since
the magnetic field in the reconnection region is very low, due to the chosen orientation of the initial background magnetic field geometry, it is easy for
electrons to get
misguided, as they are no longer tightly attached to their magnetic field lines
and therefore may encounter different electric field structures on their large
gyroradius trajectories.

\begin{figure*}[ht]
\begin{center}
 \includegraphics[width=0.6\linewidth]{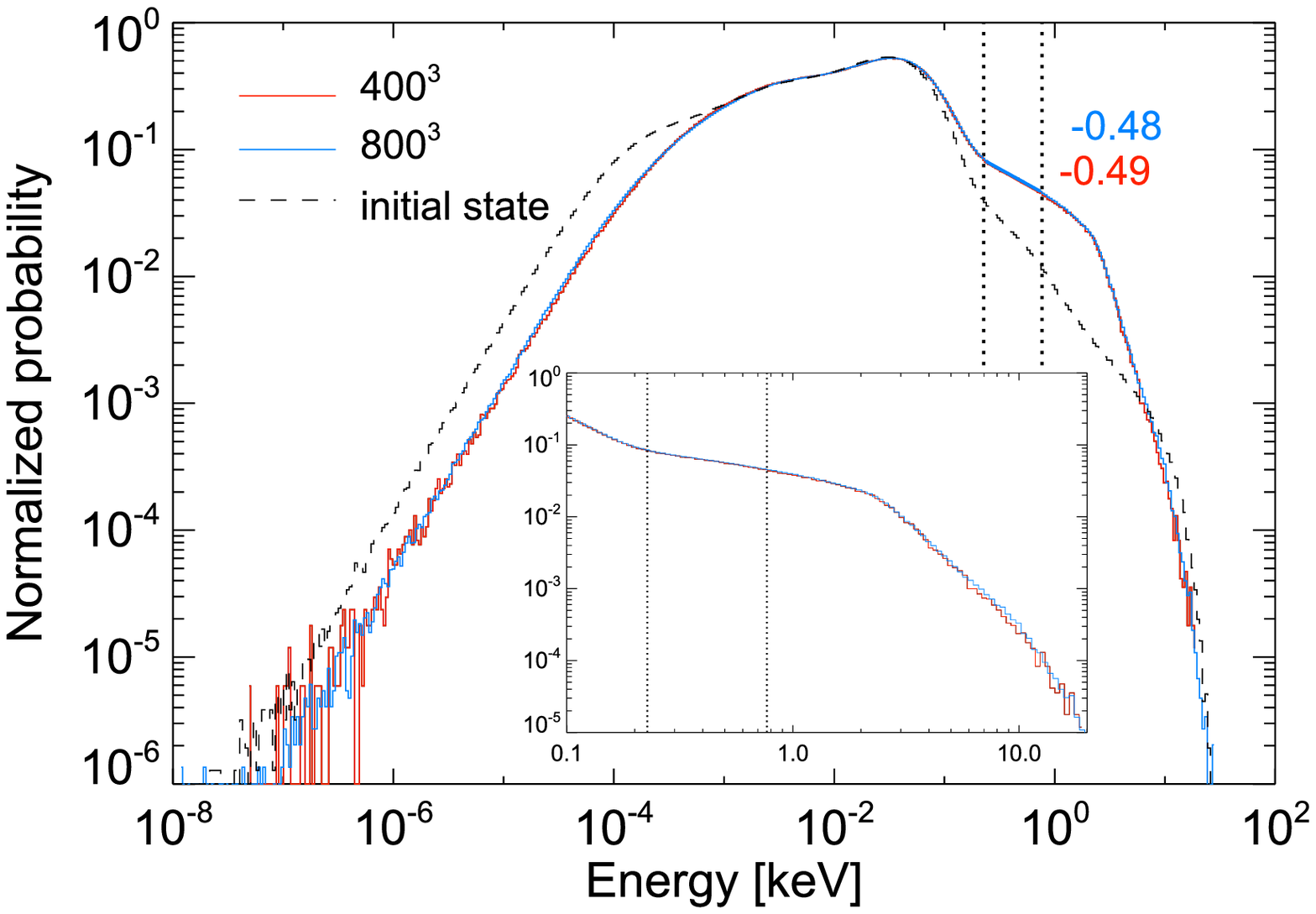}
 \includegraphics[width=0.39\linewidth]{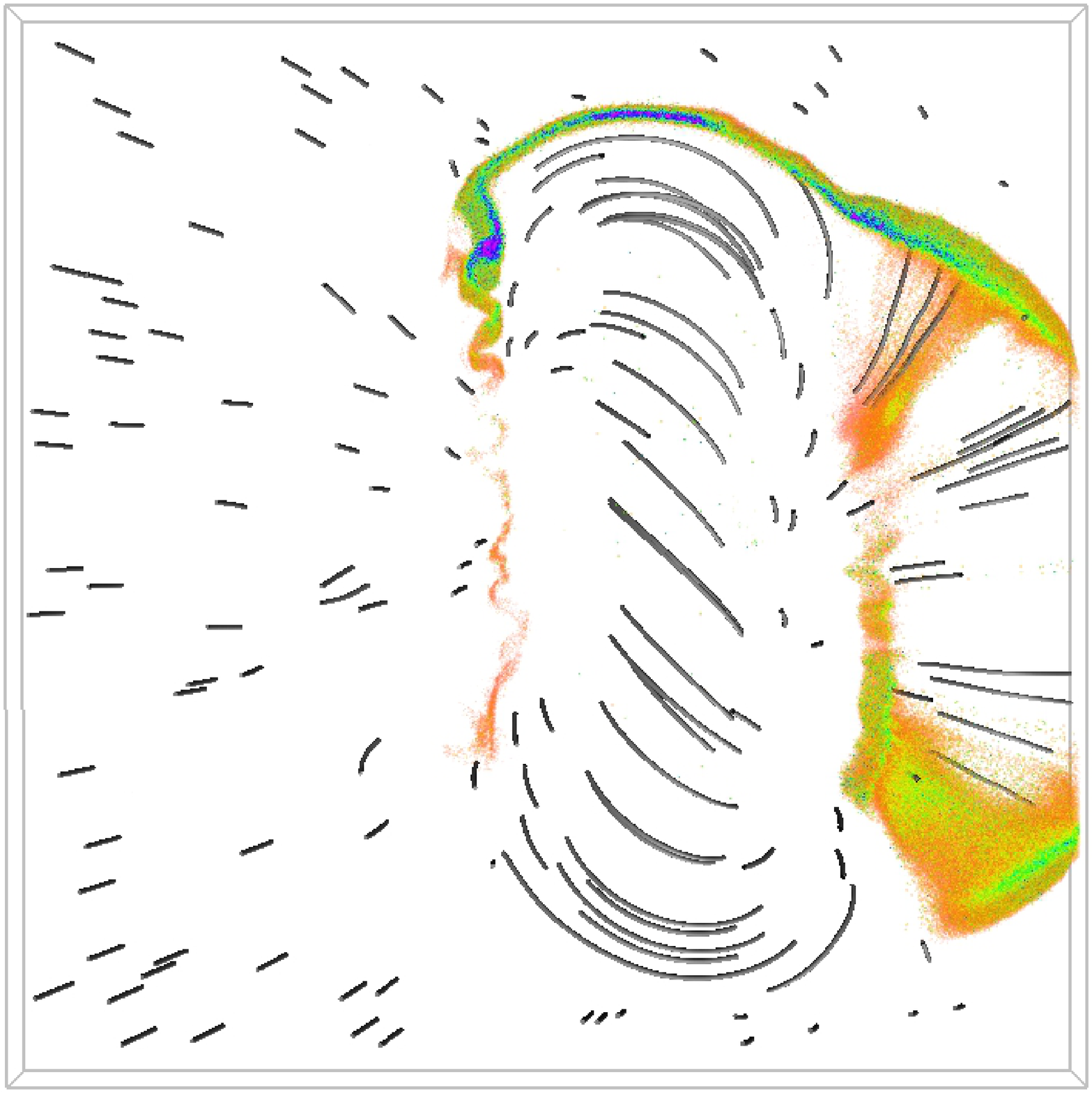}
\end{center}
  \caption{Left: normalized electron probability distribution at $t$ = 4\,s, for particles in a
  cut-out of size 19.0\,$\times$\,13.2\,$\times$\,6.5\,Mm around the reconnection region for a 400$^3$ and a comparable 800$^3$ simulation and a zoom-in. The black dashed curve illustrates
 the initial particle distribution. The vertical dotted lines mark the area for which power laws are fitted, whose indices are annotated. The lower energy limit coincides with the lowest considered electron energy for the image to the right. Right: the sum of the deposited electron energy in the lowest Mm of the computational box collected over a time period of 7\,s (red, low; blue, high). The viewing angle is as in Figure\,\ref{fig:Eres_eb_3D}, but seen from the top.}
\label{fig:energyhist}
\end{figure*}

The systematic parallel electric field building up in the current sheet area is
capable of accelerating particles to non-thermal velocities. Figure\,\ref{fig:energyhist} presents the energy histogram of electrons located in a
cut-out of 19.0\,$\times$\,13.2\,$\times$\,6.5\,Mm around the reconnection region. The initial
energy distribution is shown by the dashed line. It is primarily a superposition of the
two different plasma inflow domains of the reconnection region passed to the PIC
code through the MHD bulk velocity: one of coronal origin, hence at higher
temperatures and lower densities, and another one from plasma emerging from the
flux rope, hence at lower temperatures and higher densities. To this the drift
speed from particles in the current sheet is added, as defined by
Equation\,(\ref{equ:driftvelo}).
The distribution of probability for a particle to stay in the current sheet defines a power law in the high-energy tail of the Maxwell--Boltzmann
distribution, typically featuring an index of around --0.5 on a logarithmic scale. This is the case for electrons and protons, as the acceleration mechanism inside the current sheet is for both species the same. A series of different simulations indicates, that the influence of the physical resolution on the power-law index as well as on the upper energy limit is minuscule (see Figure\,\ref{fig:energyhist}), since the current per unit width of the current sheet is set by the magnetic field geometry. This independence also demonstrates numerical convergence. 
The tail slopes converge very rapidly toward these power-law indices, since the electron acceleration happens impulsively.
Part of the accelerated electrons escape along open magnetic field lines (see Figure\,\ref{fig:particle_pos_color.eps}) into interplanetary space and hence contribute to the solar wind. Downward moving energetic electrons impact the much denser chromosphere, which could on the Sun lead to bremsstrahlung emission. In the case of high-energy electrons, this may cause hard X-ray signatures of the type observed by RHESSI.  
Figure\,\ref{fig:energyhist} shows that most energy is deposited at the magnetic field line footpoints of the coronal loops. Observational hard X-ray signatures of solar jets described in \citet{2011ApJ...742...82K} show a three-footpoint pattern (see, e.g., event 7 in Figure\,3). Due to the choice of our flux rope being 90$^{\circ}$ inclined relative to the uniform background magnetic field, we rather get a 3D structure of footpoints, represented by three lines of electron impact regions in Figure\,\ref{fig:energyhist}. Additionally, because of the magnetic field line orientation and twist, an impact region at the intersection of the flux tube and the photosphere develops. (The impact line to the right is slightly cut because of the choice of the computational box size.)

%%%%%%%%%%%%%%%%%%%%%%%%%%%%%%%%%%%%%%%%%%%%%%%%%%%%%%%%%%%%%%%%%%%%%%%%%%%%%%%%%%%%%%
\section{\uppercase{Conclusions}}
\label{sec:conclusions}

On the basis of an MHD jet experiment, similar to the one conducted by
\citet{2008ApJ...673L.211M}, but using stretched meshes to obtain higher
spatial resolution, we have used PIC simulations
to study the acceleration of charged particles in the 3D reconnection
region of a solar coronal jet.
This is the first fully 3D kinetic model employing self-consistent fields to investigate particle acceleration in the context of solar jets. It uses a new concept of combining macroscopic with microscopic simulations.

A strong correlation is found between a slowly evolving DC electric field
located inside the current sheet and the location of the accelerated
particles (electrons and protons). The magnetic field is weak and chaotic inside the current sheet, from where most of the particles that are accelerated are quickly lost, only to be replaced by new particles, which again need to be accelerated. This represents the dissipation mechanism at work.
The systematic electric field required to constantly accelerate new particles
is, in effect, a dissipative (``resistive'', non-ideal) electric field,
sustained even though the plasma particles in this experiment are collisionless. This physical process leads to an energy probability distribution of accelerated particles, featuring a power-law tail at the high-energy side of the initial Maxwell--Boltzmann distribution, with a power-law index of about --1.5. The power-law index and the invariant energy cut-off show numerical convergence of the simulations.

The impact region of energetic electrons on the lower boundary of the box is comparable to observations of similar events and magnetic field geometries, described by \citet{2011ApJ...742...82K}.

In the future, with increasing available computational resources, we hope to
be able to resolve turbulent regions sufficiently, to enable a study of stochastic particle acceleration, expected to be able to accelerate particles to much higher energies. 

%%%%%%%%%%%%%%%%%%%%%%%%%%%%%%%%%%%%%%%%%%%%%%%%%%%%%%%%%%%%%%%%%%%%%%%%%%%%%%%%%%%%%%
\section*{\uppercase{Acknowledgments}}
\label{sec:acknowledgments}
Special thanks go to Klaus Galsgaard, Jacob Trier Frederiksen, and Troels
Haugb\o lle for very valuable discussions.
The work of G.B. was supported by the
Niels Bohr International Academy,
the SOLAIRE Research Training Network (MRTN-CT-2006-035484), and the SWIFF FP7 project (no. 263340,  http://www.swiff.eu) of the European Commission.
We acknowledge that the results have been achieved
using the PRACE and John von Neumann Institute for Computing Research
Infrastructure resource JUGENE/JUROPA based in Germany at the J\"ulich
Supercomputing Centre.
Furthermore, we acknowledge a DECI grant and grants from the Danish
Center for Scientific Computing.


\begin{thebibliography}{23}
\expandafter\ifx\csname natexlab\endcsname\relax\def\natexlab#1{#1}\fi

\bibitem[{{Archontis} {et~al.}(2005){Archontis}, {Moreno-Insertis},
  {Galsgaard}, \& {Hood}}]{2005ApJ...635.1299A}
{Archontis}, V., {Moreno-Insertis}, F., {Galsgaard}, K., \& {Hood}, A.~W. 2005,
  \apj, 635, 1299

\bibitem[{{Archontis} {et~al.}(2004){Archontis}, {Moreno-Insertis},
  {Galsgaard}, {Hood}, \& {O'Shea}}]{2004A&A...426.1047A}
{Archontis}, V., {Moreno-Insertis}, F., {Galsgaard}, K., {Hood}, A., \&
  {O'Shea}, E. 2004, \aap, 426, 1047

\bibitem[{{Baumann} {et~al.}(2012){Baumann}, {Haugb{\o}lle}, \&
  {Nordlund}}]{2012arXiv1204.4947B}
{Baumann}, G., {Haugb{\o}lle}, T., \& {Nordlund}, {\AA}. 2012, arXiv e-prints

\bibitem[{{Chifor} {et~al.}(2008){Chifor}, {Young}, {Isobe}, {Mason},
  {Tripathi}, {Hara}, \& {Yokoyama}}]{2008A&A...481L..57C}
{Chifor}, C., {Young}, P.~R., {Isobe}, H., {Mason}, H.~E., {Tripathi}, D.,
  {Hara}, H., \& {Yokoyama}, T. 2008, \aap, 481, L57

\bibitem[{{Dalla} \& {Browning}(2005)}]{2005A&A...436.1103D}
{Dalla}, S., \& {Browning}, P.~K. 2005, \aap, 436, 1103

\bibitem[{{Dalla} \& {Browning}(2006)}]{2006ApJ...640L..99D}
{Dalla}, S., \& {Browning}, P.~K. 2006, \apjl, 640, L99

\bibitem[{{Dalla} \& {Browning}(2008)}]{2008A&A...491..289D}
{Dalla}, S., \& {Browning}, P.~K. 2008, \aap, 491, 289

\bibitem[{{Drake} {et~al.}(2006){Drake}, {Swisdak}, {Che}, \&
  {Shay}}]{2006Natur.443..553D}
{Drake}, J.~F., {Swisdak}, M., {Che}, H., \& {Shay}, M.~A. 2006, \nat, 443, 553

\bibitem[{{Haugb\o lle}(2005)}]{Haugboelle:2005}
{Haugb\o lle}, T. 2005, PhD thesis, Niels Bohr Institute

\bibitem[{Haugb{\o}lle {et~al.}(2012)Haugb{\o}lle, Frederiksen, Baumann, \&
  Nordlund}]{Haugbolleetal2012}
Haugb{\o}lle, T., Frederiksen, J.~T., Baumann, G., \& Nordlund, {\AA}. 2012, in
  preparation

\bibitem[{{Hededal}(2005)}]{Hededal:2005b}
{Hededal}, C. 2005, PhD thesis, Niels Bohr Institute 

\bibitem[{{Kamio} {et~al.}(2007){Kamio}, {Hara}, {Watanabe}, {Matsuzaki},
  {Shibata}, {Culhane}, \& {Warren}}]{2007PASJ...59S.757K}
{Kamio}, S., {Hara}, H., {Watanabe}, T., {Matsuzaki}, K., {Shibata}, K.,
  {Culhane}, L., \& {Warren}, H.~P. 2007, \pasj, 59, 757

\bibitem[{{Krucker} {et~al.}(2011){Krucker}, {Kontar}, {Christe}, {Glesener},
  \& {Lin}}]{2011ApJ...742...82K}
{Krucker}, S., {Kontar}, E.~P., {Christe}, S., {Glesener}, L., \& {Lin}, R.~P.
  2011, \apj, 742, 82

\bibitem[{{Krucker} {et~al.}(2007){Krucker}, {Kontar}, {Christe}, \&
  {Lin}}]{2007ApJ...663L.109K}
{Krucker}, S., {Kontar}, E.~P., {Christe}, S., \& {Lin}, R.~P. 2007, \apjl,
  663, L109  

\bibitem[{{Moreno-Insertis} {et~al.}(2008){Moreno-Insertis}, {Galsgaard}, \&
  {Ugarte-Urra}}]{2008ApJ...673L.211M}
{Moreno-Insertis}, F., {Galsgaard}, K., \& {Ugarte-Urra}, I. 2008, \apjl, 673,
  L211

\bibitem[{{Rosdahl} \& {Galsgaard}(2010)}]{2010A&A...511A..73R}
{Rosdahl}, K.~J., \& {Galsgaard}, K. 2010, \aap, 511, A73

\bibitem[{{Savcheva} {et~al.}(2007){Savcheva}, {Cirtain}, {Deluca},
  {Lundquist}, {Golub}, {Weber}, {Shimojo}, {Shibasaki}, {Sakao}, {Narukage},
  {Tsuneta}, \& {Kano}}]{2007PASJ...59S.771S}
{Savcheva}, A., {Cirtain}, J., {Deluca}, E.~E., {Lundquist}, L.~L., {Golub},
  L., {Weber}, M., {Shimojo}, M., {Shibasaki}, K., {Sakao}, T., {Narukage}, N.,
  {Tsuneta}, S., \& {Kano}, R. 2007, \pasj, 59, 771

\bibitem[{{Siversky} \& {Zharkova}(2009)}]{2009JPlPh..75..619S}
{Siversky}, T.~V., \& {Zharkova}, V.~V. 2009, J. of Plasma Phys., 75,
  619

\bibitem[{{Turkmani} {et~al.}(2006){Turkmani}, {Cargill}, {Galsgaard},
  {Vlahos}, \& {Isliker}}]{2006A&A...449..749T}
{Turkmani}, R., {Cargill}, P.~J., {Galsgaard}, K., {Vlahos}, L., \& {Isliker},
  H. 2006, \aap, 449, 749

\bibitem[{{Turkmani} {et~al.}(2005){Turkmani}, {Vlahos}, {Galsgaard},
  {Cargill}, \& {Isliker}}]{2005ApJ...620L..59T}
{Turkmani}, R., {Vlahos}, L., {Galsgaard}, K., {Cargill}, P.~J., \& {Isliker},
  H. 2005, \apjl, 620, L59

\end{thebibliography}
\end{document}